\documentclass[12pt]{article}
\usepackage{graphicx,amsmath,amssymb}
\usepackage{indentfirst}
\usepackage[usenames]{color}
\usepackage[colorlinks=true, urlcolor=navyblue, linkcolor=navyblue, citecolor=navyblue]{hyperref}
\usepackage{epstopdf}
\usepackage{enumerate}
\usepackage{appendix}
\usepackage{lineno}
\usepackage{setspace}

\definecolor{navyblue}{rgb}{0,0.08,0.45}
\definecolor{darkred}{rgb}{0.7,0.0,0.0}
\definecolor{darkgreen}{rgb}{0,0.6,0.2}

\newcommand{\beq}{\begin{equation}}
\newcommand{\enq}{\end{equation}}
\newcommand{\beqa}{\begin{eqnarray}}
\newcommand{\beqast}{\begin{eqnarray*}}
\newcommand{\enqa}{\end{eqnarray}}
\newcommand{\enqast}{\end{eqnarray*}}

\newcommand{\bec}{\begin{center}}
\newcommand{\enc}{\end{center}}
\newcommand{\beqo}{\begin{quote}}
\newcommand{\enqo}{\end{quote}}
\newcommand{\bem}{\begin{minipage}}
\newcommand{\enm}{\end{minipage}}

\begin{document}

\vspace{15pt}

\begin{center}
{\huge  Light-Front Holography, Color Confinement, }

\vspace{10pt}

{\huge    and Supersymmetric Features of QCD}

\end{center}

\vspace{15pt}

\centerline{Stanley J. Brodsky}

\vspace{3pt}

\centerline {\it SLAC National Accelerator Laboratory, Stanford University~\footnote{{Invited talk presented at Light Cone 2015,  
September 21-25, 2015, INFN, Frascati, Italy\\
\href{mailto:sjbth@slac.stanford.edu}{\tt
\hspace{12pt} sjbth@slac.stanford.edu}}}}

\vspace{20pt}

\begin{abstract}

Light-Front Quantization --  Dirac's ``Front Form" -- provides a physical, frame-independent formalism for hadron dynamics and structure.  Observables such as structure functions, transverse momentum distributions, and distribution amplitudes are defined from the hadronic light-front wavefunctions.   One obtains new insights into the hadronic 
spectrum, light-front wavefunctions, and the functional form of the QCD running coupling in the nonperturbative domain using light-front holography -- the duality between the front form   and AdS$_5$, the space of isometries of the conformal group. 
In addition, superconformal algebra leads to remarkable supersymmetric relations between mesons and baryons of the same parity.  The mass scale $\kappa$ underlying confinement and hadron masses  can be connected to the parameter   $\Lambda_{\overline {MS}}$ in the QCD running coupling by matching the nonperturbative dynamics, as described by  the effective conformal theory mapped to the light-front and its embedding in AdS space, to the perturbative QCD  regime. The result is an effective coupling  defined at all momenta.   This 
matching of the high and low momentum transfer regimes determines a scale $Q_0$ which  sets the interface between perturbative and nonperturbative hadron dynamics.  
The use of $Q_0$ to  resolve  the factorization scale uncertainty for structure functions and distribution amplitudes,  in combination with the principle of maximal conformality (PMC)  for  setting the  renormalization scales,  can 
greatly improve the precision of perturbative QCD predictions for collider phenomenology.  The absence of vacuum excitations of the causal, frame-independent  front form vacuum has important consequences  for the cosmological constant.   I also discuss evidence that the antishadowing of nuclear structure functions is non-universal; {\it i.e.},  flavor dependent, and why shadowing and antishadowing phenomena may be incompatible with the momentum and other sum rules for nuclear parton distribution functions.     \end{abstract}

\newpage

\tableofcontents

\section{Introduction}

Light-front quantization provides a physical, frame-independent formalism for hadron dynamics and structure.  
When one makes a measurement of a hadron, such as in deep inelastic lepton-proton scattering $\ell p \to \ell^\prime X$, the hadron is observed
along a light-front (LF) --  in analogy to a flash photograph -- not at a fixed time $t$.  In effect, the LF  time variable  is 
$\tau= x^+ = t+z/c$, the time tangent to the light-cone.  This is the underlying principle of the ``front form" postulated  by Dirac~\cite{Dirac:1949cp}.
 
The front form has the maximum number of kinematic generators of the Lorentz group, and most remarkably, the formalism is boost invariant.
The LF time evolution operator $P^- \equiv P^0-P^z = i  {d\over d\tau}$ and the corresponding LF Hamiltonian   $H_{LF} =  P^+ P^- -{\vec P}^2_\perp$, where $P^+=P^0+P^z$ and $\vec P_\perp$ are kinematical, can be derived directly from the Lagrangian.  
In the  case of QCD, the eigenvalues of the LF invariant Hamiltonian are the squares of the hadron masses $M^2_H$: 
$H_{LF}|\Psi_H>  = M^2_H |\Psi_H>$~\cite{Brodsky:1997de}, and the corresponding eigensolutions provide the $n$-particle hadronic LF Fock state wavefunctions (LFWFs)
 $<n| \Psi_H> = \psi^H_n(x_i, \vec k_{\perp i },\lambda_i)$,  where $|n>$ projects the eigenstate on the free Fock basis.  The constituents' physical momenta are 
$p^+_i = x_i P^+$, and  $\vec p_{\perp i } =  x_i  {\vec P}_\perp +  \vec k_{\perp i }$,  and the $\lambda_i$ label the  spin projections $S^z_i$.
One can avoid ghosts and longitudinal  gluonic degrees of freedom by choosing to work in the light-cone gauge  $A^+ =0$.  

The LFWFs are thus the Fock state projections of the eigenstates of the QCD LF Hamiltonian.
The LFWFs are boost invariant; {\it i.e.}, independent of the hadron's longitudinal  $P^+ =P^0 +P^z$ and transverse momentum $ \vec P_\perp.$
Observables such as hadron structure functions, form factors, transverse momentum distributions, weak-decay amplitudes and distribution amplitudes are defined directly from the hadronic LFWFs.  An example of a calculation of deeply virtual Compton scattering is given in ref. \cite{Brodsky:2000xy}.
  
\begin{figure}
\begin{center}
\includegraphics[height=10cm,width=15cm]{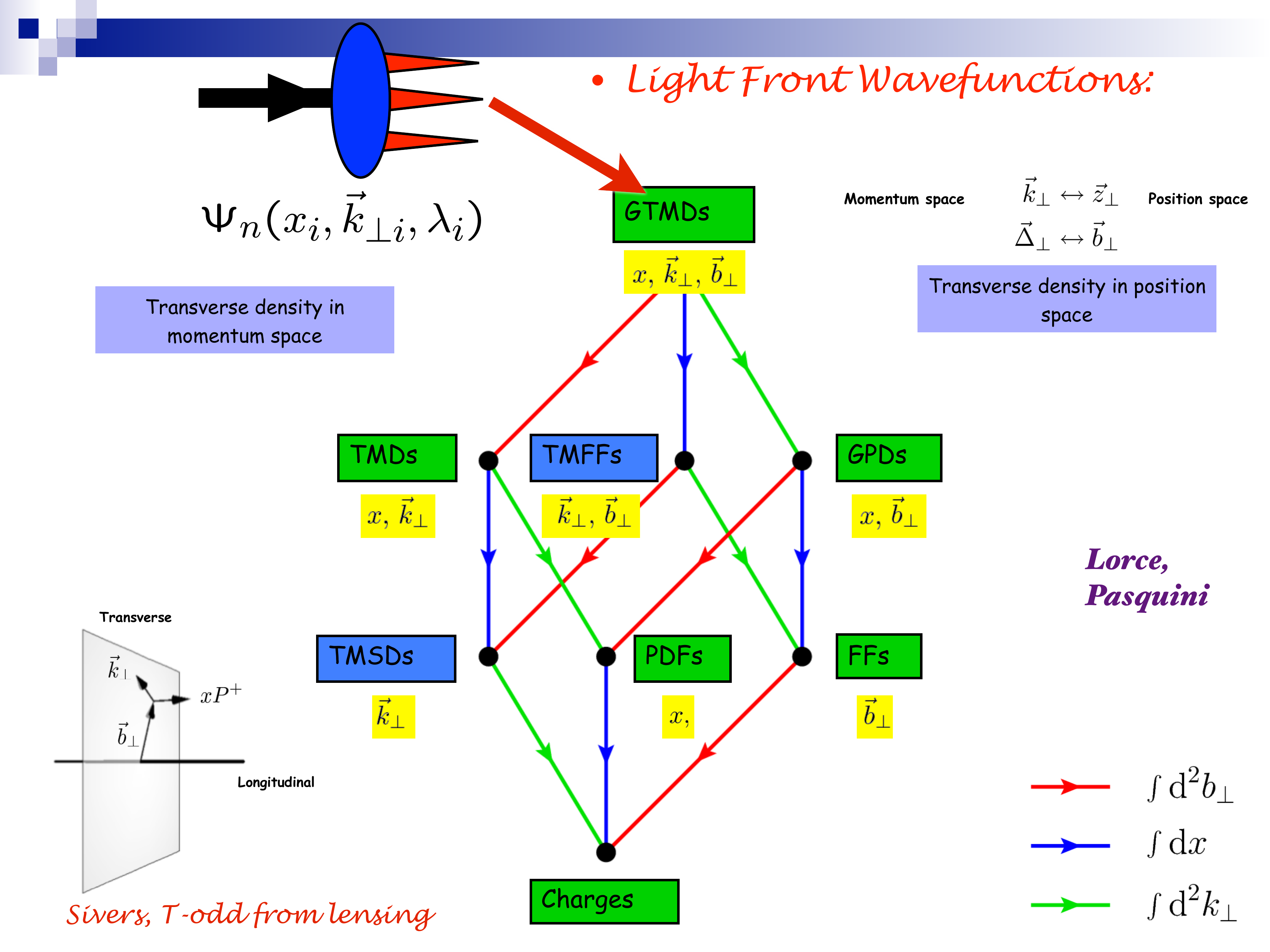}
\end{center}
\caption{Measures such as the generalized transverse momentum distribution (GTMD) which are directly defined from overlaps of hadronic light-front wavefunctions. In some cases, such as the leading twist  Sivers  pseudo-T-odd spin  correlation in polarized deep inelastic lepton-proton scattering~\cite{Brodsky:2002cx} or diffractive deep inelastic scattering $\ell p \to \ell^\prime p^\prime X$~\cite{Brodsky:2002ue}, one must include final-state ``lensing"  interactions.  Adopted from a figure designed by B. Pasquini and C. Lorce~\cite{Pasquini:2014fja}.
}
\label{LFMeasures.pdf}
\end{figure}

The LF analyses are causal and frame independent.   Operators appearing in commutators are automatically normal-ordered since there are no quantum fluctuations created from the LF vacuum. The LF wavefunctions thus play the role of Schr\"odinger wavefunctions in atomic physics, but they are relativistic and frame independent,
see Fig. \ref{LFMeasures.pdf}.
Since they are independent of $P^+=P^0+ P^z$ and $\vec P_\perp$, the LFWFs are the same whether measured in the hadron rest frame, as in the SLAC experiments or at a hadron collider, such as at an electron-proton collider. There is thus no concept of length contraction of a moving hadron or nucleus in the front form since observations  of the collisions of the composite hadrons are not made  at  fixed instant  time $t$.    
The absence of length contraction of a photographed object was first noted by Terrell~\cite{Terrell:1959zz}, Penrose~\cite{Penrose:1959vz}, and Weisskopf ~\cite{Weisskopf}. 
LF quantization is thus the natural formalism for particle physics.  

There is an effort in the lattice gauge theory community to boost LGTh results to infinite momentum in order to simulate the causal light-front results~\cite{Ma:2014jla}. 
An alternative lattice method -- the `` transverse lattice"   introduces  light-front coordinates in an effective 1+1 gauge theory connected to a lattice in the remaining  transverse dimensions~\cite{Bardeen:1978gw,Bardeen:1979xx,Burkardt:2001jg}.

A review of the light-front formalism is given in Ref.~\cite{Brodsky:1997de}.

A key result  for hadron physics is the Drell-Yan-West  formula~\cite{Drell:1969km, West:1970av,Brodsky:1980zm}  for electromagnetic form factors measured in elastic lepton hadron scattering.  
If one chooses the special frame  $q^+= \ell^+ - \ell^{\prime +} =0$, then the current matrix element $<p+q|j^+(0)|p>$, which defines form factors, is a simple overlap integral of the initial $\psi_H(x_i, \vec k_{\perp i },\lambda_i)$ and final-state  LFWFs $ \psi_H(x_i, \vec k^\prime_{\perp i }, \lambda_i)$, summed over Fock states $n$.  This yields the exact results for the electromagnetic, weak, and gravitational form factors.  Here $\vec k^\prime_{\perp i } = \vec k^\prime_{\perp i } + (1-x_i) \vec q_\perp$ for the struck quark and 
$\vec k^\prime_{\perp i } = \vec k^\prime_{\perp i }  - x_i \vec q_\perp$ for the spectators.   For example, one can easy show that the anomalous gravitomagnetic moment $B(q^2=0)$ vanishes identically for any LF Fock state~\cite{Brodsky:2000ii}, in agreement with a theorem by Okun, Kobzarev, and Teryaev~\cite{Kobzarev:1962wt,Teryaev:1999su} which follows from the equivalence theorem of gravity.

The LF predictions for current matrix elements are equivalent to the corresponding Bethe-Salpeter formulae for form factors, but one must sum over the currents of an infinite number of irreducible kernels.  The LFWFs are related to the Bethe Salpeter amplitudes if one integrates over $k^-$. 

The LFWFs are boost invariant.
This contrasts with the wavefunctions defined at a fixed time $t$ -- the Lorentz boost of an instant-form wavefunction is a dynamical problem, much more complicated than a simple
Melosh transform~\cite{Brodsky:1968ea} -- even the number of Fock constituents changes under a boost.    Thus there is no practical formula for calculating form factors using ordinary instant time  analogous to the Drell-Yan West formula since the boost of the wavefunction of a composite state is unknown except for weakly bound systems;  and worse, one must include the contributions of an infinite number of connected currents which arise from the vacuum.  

PQCD factorization theorems and  the DGLAP  \cite{Gribov:1972ri,Altarelli:1977zs,Dokshitzer:1977sg} and ERBL \cite{Lepage:1979zb,Lepage:1980fj,Efremov:1979qk,Efremov:1978rn} evolution equations can also be derived using the light-front Hamiltonian formalism~\cite{Lepage:1980fj}.  In the case of an electron-ion collider, one can represent the cross section for $e-p$ colisions as a convolution of the hadron and virtual photon structure functions times the subprocess cross-section in analogy to hadron-hadron colisions.   This nonstandard description of $\gamma^* p \to X$ reactions  gives new insights into electroproduction physics -- physics not apparent   in the usual infinite-momentum frame description, such as the dynamics of heavy quark-pair production.  
Intrinsic heavy quarks at high $x$  also play an important role~\cite{Brodsky:2015uwa}.

The LF Heisenberg equation can in principle be solved numerically by matrix diagonalization  using the ``Discretized Light-Cone  Quantization" (DLCQ)~\cite{Pauli:1985pv} method.  Anti-periodic boundary conditions in 
$x^-$ render the $k^+$ momenta  discrete  as well as  limiting the size of the Fock basis.   In fact, one can easily solve $1+1 $ quantum field theories such as QCD$(1+1)$~\cite{Hornbostel:1988fb} for any number of colors, flavors and quark masses using DLCQ. 
Unlike lattice gauge theory, the nonpertubative DLCQ analysis is in Minkowski space, is frame-independent and is free of fermion-doubling problems.

As I shall discuss, AdS/QCD, together with light-front holography, is now providing a  color-confining approximation to $H_{LF}^{QCD}$,  for QCD(3+1), thus giving a first approximation to the meson and baryon spectra and their hadronic LFWFs.  A new method for solving nonperturbative QCD ``Basis Light-Front Quantization" (BLFQ)~\cite{Vary:2014tqa},  uses the eigensolutions of a color-confining approximation to QCD (such as LF holography) as the basis functions,  rather than the plane-wave basis used in DLCQ.  LFWFs can also be determined from the covariant Bethe-Salpeter wavefunction by integrating over $k^-$~\cite{Brodsky:2015aia}.  
There is also now an effort in the lattice gauge theory community to boost LGTh results to infinite momentum to simulate the causal light-front results.

\section{Calculations using LF-Time-Ordered Perturbation Theory}

There are a number of advantages if one uses  LF Hamiltonian methods for perturbative QCD calculations.   

Propagating particles  are on their respective mass shells:  $k_\mu k^\mu = m^2$, and intermediate states are off-shell in invariant mass;  {\it i.e.}, $P^- \ne \sum k^-_i$.
Unlike instant form, where one must sum  $n !$ frame-dependent  amplitudes, only  $\tau$-ordered diagrams where every line has  positive $k^+ =k^0+k^z$  contribute.  An excellent example  of LF-time-ordered perturbation theory is the computation of multi-gluon scattering amplitudes by Cruz-Santiago and Stasto~\cite{Cruz-Santiago:2015dla}.  The number of nonzero amplitudes is also greatly reduced by noting that the total angular momentum projection $J^z = \sum_i^{n-1 } L^z_i + \sum^n_i S^z_i$ and the total $P^+$ are  conserved at each vertex.  In addition, in a renormalizable theory the change in orbital angular momentum is limited to $\Delta L^z =0,\pm 1$ at each vertex.  

A remarkable advantage of LF time-ordered perturbation theory (LFPth) is that the calculation of a subgraph of any order in pQCD only needs to be done once;  the result can be stored in a ``history" file.  This is due to the fact that in LFPth the numerator algebra is independent of the process; the denominator changes, but only by a simple shift of the initial $P^-$.   Another simplification is that loop integrations are three dimensional: $\int d^2\vec k_\perp \int^1_0 dx.$   Unitarity  and explicit
 renormalization can be implemented using the ``alternate denominator" method which defines the required subtraction counterterms~\cite{Brodsky:1973kb}.

\section{The Light-Front Vacuum}

It is important to distinguish the LF vacuum from the conventional instant-form vacuum.

The eigenstates of the instant-form Hamiltonian describe a state defined at a single instant of time $t$ over all space, and they are thus acausal as well as frame-dependent.  
The instant-form vacuum is defined as the lowest energy eigenstate of the instant-form Hamiltonian.
Quantum loops in the  instant-form vacuum  typically vanish as $1/ P^2$.

In contrast, the vacuum in LF Hamlitonian theory is defined as the eigenstate of $H_{LF}$ with lowest invariant mass.  It is defined at fixed {color{blue} LF time} $\tau$ within the causal horizon and is frame-independent. 
The loop diagrams which occur in the usual instant-form vacuum do not appear  in the front-form vacuum since  the  $+$ momenta are positive: $k^+ _i = k^0_i+k^z_i\ge 0$, and the sum of $+$ momenta is conserved at every vertex.  Thus the creation of particles cannot arise from the LF vacuum  since $ \sum_i  k^{+i}   \ne P^+_{vacuum} =0.$  
Since propagation with negative $k^+$  does not appear, the LF vacuum is trivial up to possible $k^+=0$ ``zero"  modes.   The usual quark and gluon QCD vacuum condensates of the instant form are replaced by physical effects contained within the hadronic LFWFs  in the hadronic domain. 
This is referred to as ``in-hadron" condensates~\cite{Casher:1974xd,Brodsky:2009zd,Brodsky:2010xf}.  In the case of the Higgs theory, the traditional Higgs vacuum expectation value (VEV) is replaced by a ``zero mode", analogous to a classical 
Stark or Zeeman field~\cite{Srivastava:2002mw}.   This again contrasts with the traditional view of the vacuum  based on the instant form. 

The cosmological constant  is of order $10^{120} $ times larger than what is observed if one computes the effects of quantum loops from QED using the instant form vacuum.  QCD instantons and condensates in the same vacuum give a contribution of order  $10^{42}.$  The contribution of the  Higgs VEV computed in the instant form vacuum is  $10^{54}$ times too large~\cite{Zee:2008zz}.  Conventional wisdom~\cite{Bailey} suggests that this problem can be solved by assuming the existence of  $10^{500} $ possible universes, fine-tuning to 120 decimal places,  together with the anthropic principle.

However, the universe is observed within the causal horizon, not at a single instant of time.  The causal, frame-independent light-front vacuum can thus provide a viable match to the empty visible universe~\cite{Brodsky:2010xf}.  The cosmological constant  problem thus does not appear if one notes that the causal, frame-independent light-front vacuum has no quantum fluctuations --  in dramatic contrast to to  the acausal, frame-dependent instant-form vacuum;  the cosmological constant vanishes if one uses the front form.     

In the case of electroweak theory, the Higgs LF zero mode~\cite{Srivastava:2002mw}  has no energy-momentum density,  so it also gives zero contribution to the cosmological constant.  However, it is possible that if one solves electroweak theory in a curved universe, the Higgs LF zero mode will be replaced with a field of nonzero curvature which could give a nonzero contribution.

\section{Color Confinement and Supersymmetry in Hadron Physics from LF Holography}

\begin{figure}
 \begin{center}
\includegraphics[height= 12cm,width=15cm]{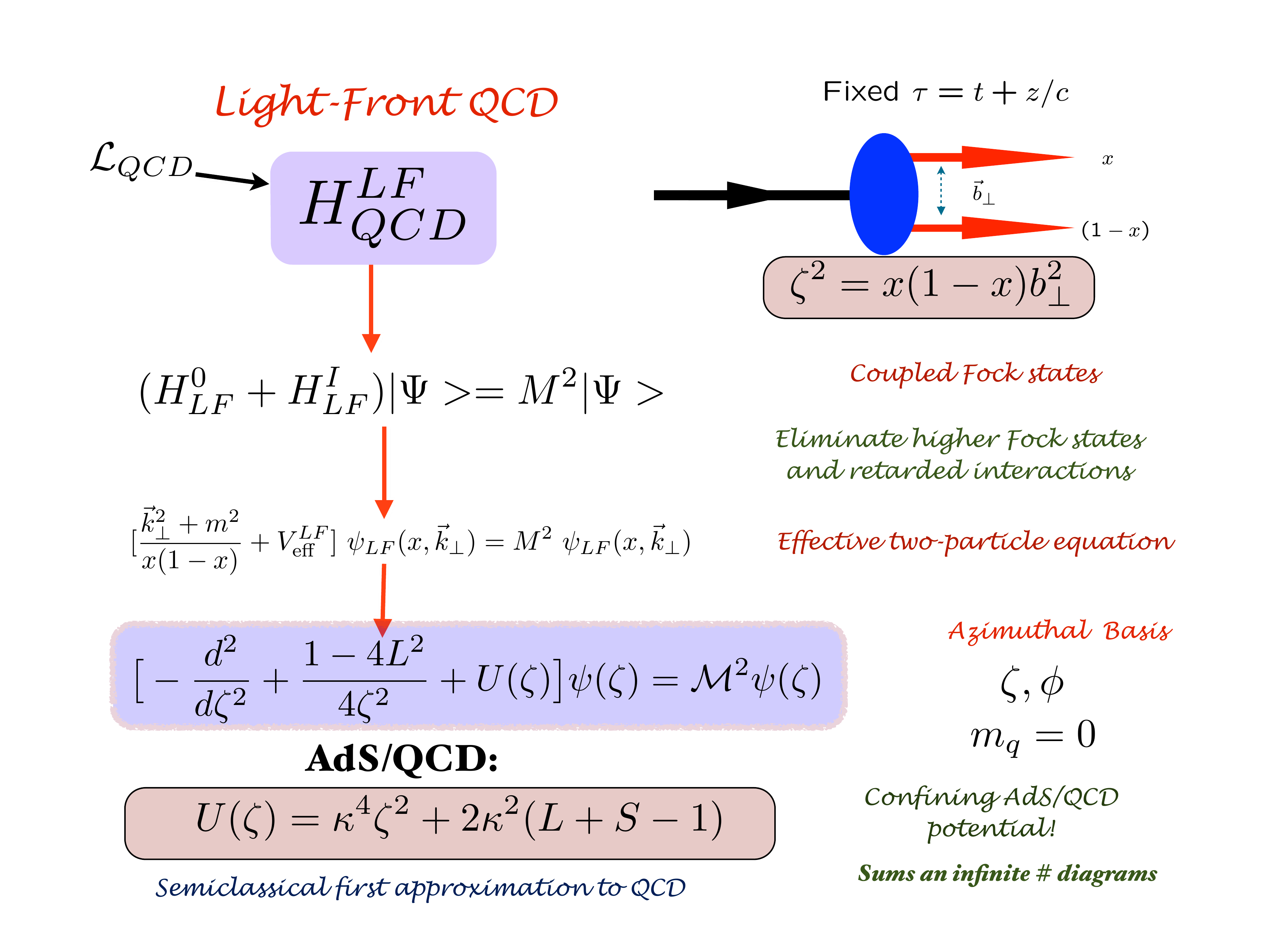}
\end{center}
\caption{Derivation of the Effective Light-Front Schr\"odinger Equation from QCD.  As in QED, one reduces the LF Heisenberg equation $H_{LF}|\Psi >   = M^2 |\Psi>$ 
to an effective two-body eigenvalue equation for $q \bar q$ mesons by systematically eliminating higher Fock states. One utilizes the LF radial variable $\zeta$, where $\zeta^2 = x(1-x)b^2_\perp$ is conjugate to the $q \bar q$ LF kinetic energy $k^2_\perp\over x(1-x)$ for $m_q=0$. This allows the reduction of the dynamics to a single-variable bound state equation acting on the valence $q \bar q$ Fock state.  The confining potential $U(\zeta)$, including its spin-$J$ dependence, is derived from the soft-wall AdS/QCD model with the dilaton  $e^{+\kappa^2 z^2 },$ where $z$ is the fifth coordinate of AdS$_5$ holographically dual  to $\zeta$. See Ref.~\cite{Brodsky:2013ar}.   The resulting light-front harmonic oscillator confinement potential $\kappa^4 \zeta^2 $ for light quarks is equivalent to a linear confining potential for heavy quarks in the instant form~\cite{Trawinski:2014msa}. }
\label{FigsJlabProcFig2.pdf}
\end{figure}

A key problem in hadron physics is to obtain a first color-confining approximation to QCD which can predict both the hadron spectrum and the hadronic LFWFs.  
If one neglects the Higgs couplings of quarks, then no mass parameter appears in the QCD Lagrangian, and the theory is conformal at the classical level.   Nevertheless,  hadrons have a finite mass.  De T\'eramond, Dosch, and I~\cite{Brodsky:2013ar}
have shown that a mass gap and a fundamental color confinement scale can be derived from a conformal action when one extends the formalism of de Alfaro, Fubini and Furlan  (dAFF)~\cite{deAlfaro:1976je}  to light-front Hamiltonian theory. Remarkably, the resulting light-front potential has a unique form of a harmonic oscillator $\kappa^4 \zeta^2$ in the 
light-front invariant impact variable $\zeta$ where $ \zeta^2Ê = b^2_\perp x(1-x)$. The result is  a single-variable frame-independent relativistic equation of motion for  $q \bar q $ bound states, a ``Light-Front Schr\"odinger Equation"~\cite{deTeramond:2008ht}, analogous to the nonrelativistic radial Schr\"odinger equation in quantum mechanics.  The  Light-Front Schr\"odinger Equation  incorporates color confinement and other essential spectroscopic and dynamical features of hadron physics, including a massless pion for zero quark mass and linear Regge trajectories with the same slope  in the radial quantum number $n$   and internal  orbital angular momentum $L$.   
The same light-front  equation for mesons of arbitrary spin $J$ can be derived~\cite{deTeramond:2013it}
from the holographic mapping of  the ``soft-wall model" modification of AdS$_5$ space with the specific dilaton profile $e^{+\kappa^2 z^2},$  where one identifies the fifth dimension coordinate $z$ with the light-front coordinate $\zeta$.  The five-dimensional AdS$_5$ space provides a geometrical representation of the conformal group.
It is holographically dual to 3+1  spacetime at fixed light-front time $\tau = t+ z/c$.  
The derivation of the confining LF Schrodinger Equation is outlined in Fig. \ref{FigsJlabProcFig2.pdf}. 
The reduction to an effective Hamiltonian acting on the valence Fock state of hadrons in QCD is analogous to the reduction used in precision analyses in QED for atomic physics.

\begin{figure}
 \begin{center}
\includegraphics[height= 12cm,width=15cm]{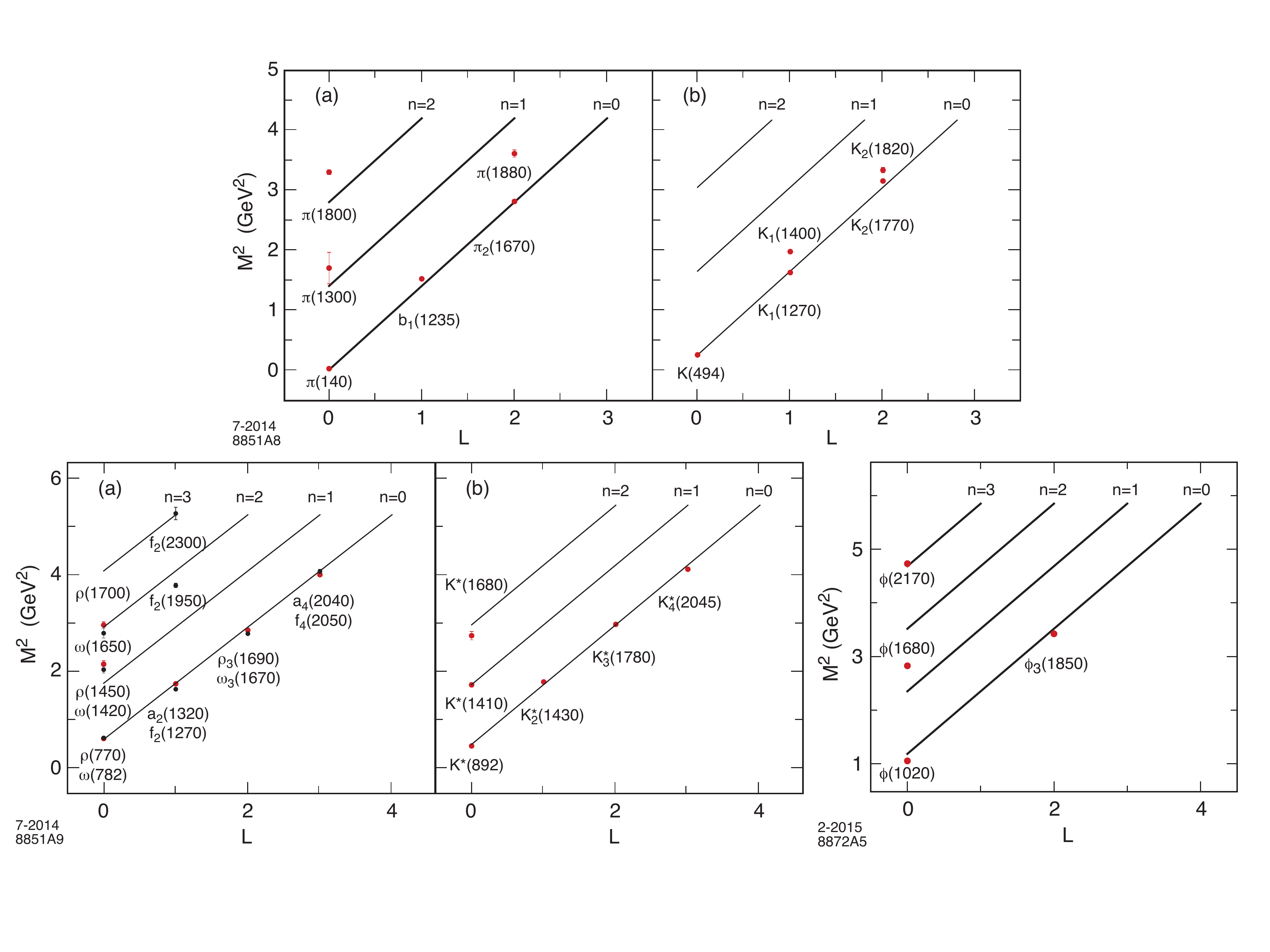}
\end{center}
\caption{Comparison of the AdS/QCD prediction  $M^2(n, L, S) = 4\kappa^2(n+L +S/2)$ for the orbital $L$ and radial $n$ excitations of the meson spectrum with experiment.   The pion is predicted to be massless for zero quark mass. The $u,d,s$ quark masses can be taken into account by perturbing in $<m_q^2/x>$.   The fitted value of $\kappa = 0.59$ MeV for pseudoscalar mesons, 
and  $\kappa = 0.54$ MeV  for vector mesons. }
\label{ReggePlot.pdf}
\end{figure}

The  combination of light-front dynamics, its holographic mapping to AdS$_5$ space, and the dAFF procedure provides new  insight into the physics underlying color confinement, the nonperturbative QCD coupling, and the QCD mass scale.  A comprehensive review is given in  Ref.~\cite{Brodsky:2014yha}.  The $q \bar q$ mesons and their valence LF wavefunctions are the eigensolutions of the frame-independent relativistic bound state LF Schr\"odinger equation.  The mesonic $q\bar  q$ bound-state eigenvalues for massless quarks are $M^2(n, L, S) = 4\kappa^2(n+L +S/2)$.
The equation predicts that the pion eigenstate  $n=L=S=0$ is massless at zero quark mass. The  Regge spectra of the pseudoscalar $S=0$  and vector $S=1$  mesons  are 
predicted correctly, with equal slope in the principal quantum number $n$ and the internal orbital angular momentum $L$.  The comparison with experiment is shown in Fig. \ref{ReggePlot.pdf}.

\begin{figure}
\begin{center}
\includegraphics[height=10cm,width=15cm]{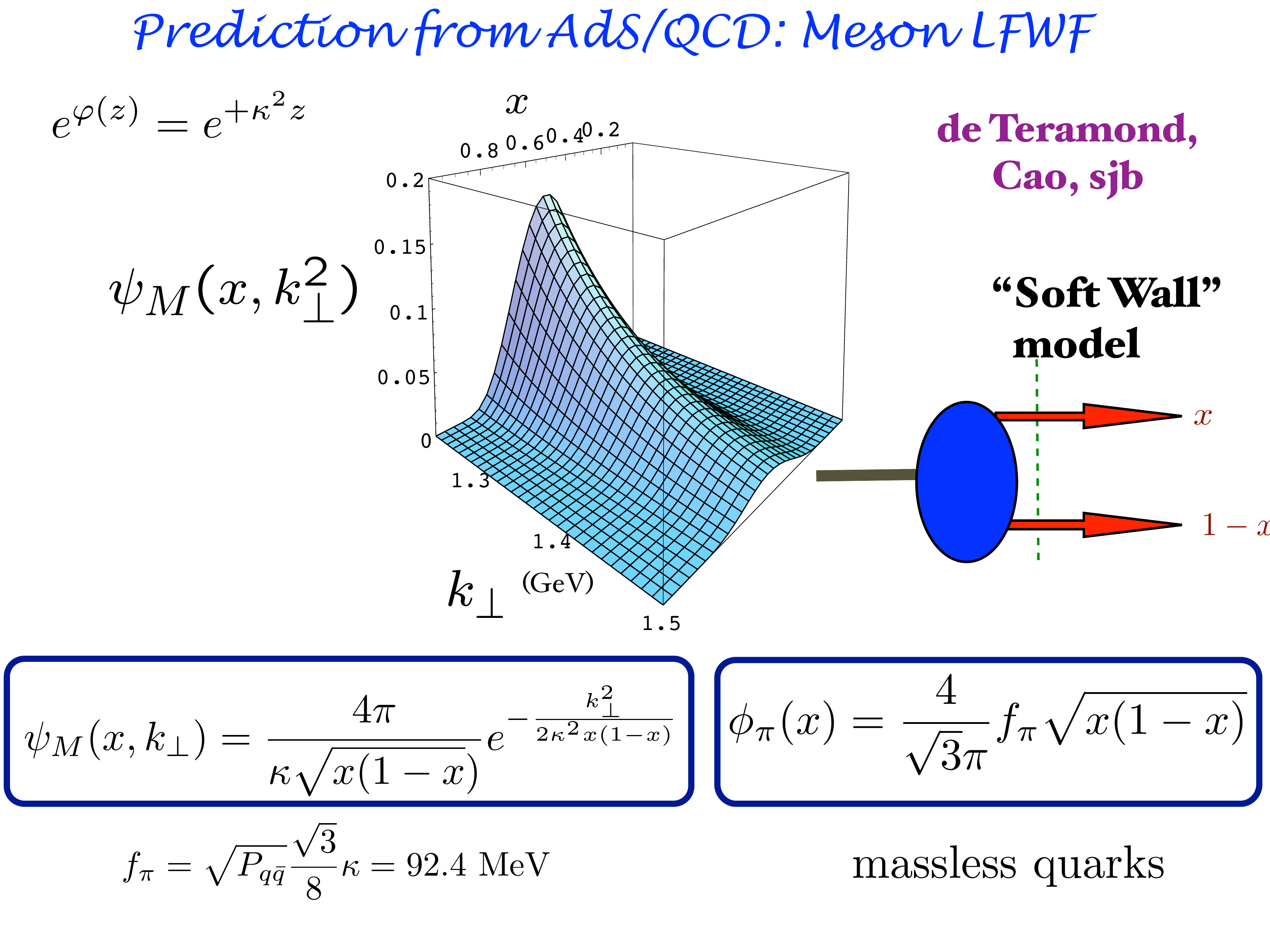}
\end{center}
\caption{Prediction from AdS/QCD and Light-Front Holography for  meson LFWFs  $\psi_M(x,\vec k_\perp)$   and the pion distribution amplitude.     
}
\label{MesonLFWF.pdf}
\end{figure}

The AdS/QCD light-front holographic eigenfunction for the $\rho$ meson LFWF $\psi_\rho(x,\vec k_\perp)$ gives excellent 
predictions for the observed features of diffractive $\rho$ electroproduction $\gamma^* p \to \rho  p^\prime$,  as shown by Forshaw and Sandapen~\cite{Forshaw:2012im}.  
Note that the prediction for the LFWF is a function of the LF kinetic energy $\vec k^2_\perp/ x(1-x)$ -- the conjugate of the LF radial variable $\zeta^2 = b^2_\perp x(1-x)$ -- times a function of $x(1-x)$. It does not factorize as a  function of $\vec k^2_\perp$ times a function of $x$.  The resulting  nonperturbative pion distribution amplitude $\phi_\pi(x) = \int d^2 \vec k_\perp \psi_\pi(x,\vec k_\perp) = (4/  \sqrt 3 \pi)  f_\pi \sqrt{x(1-x)}$,  see Fig. \ref{MesonLFWF.pdf}, is  consistent with the Belle data for the photon-to-pion transition form factor~\cite{Brodsky:2011xx}. 

\section{Supersymmetric Aspects of Hadron Physics}
These results can be extended~\cite{deTeramond:2014asa,Dosch:2015nwa,Dosch:2015bca} to effective QCD light-front equations for both mesons and baryons by using the generalized supercharges of superconformal algebra~\cite{Fubini:1984hf}.   In effect the baryons are color-singlet bound-states of color-triplet quarks and $\bar 3_C$ $[qq]$ diquarks.  

\begin{figure}
 \begin{center}
\includegraphics[height=10cm,width=15cm]{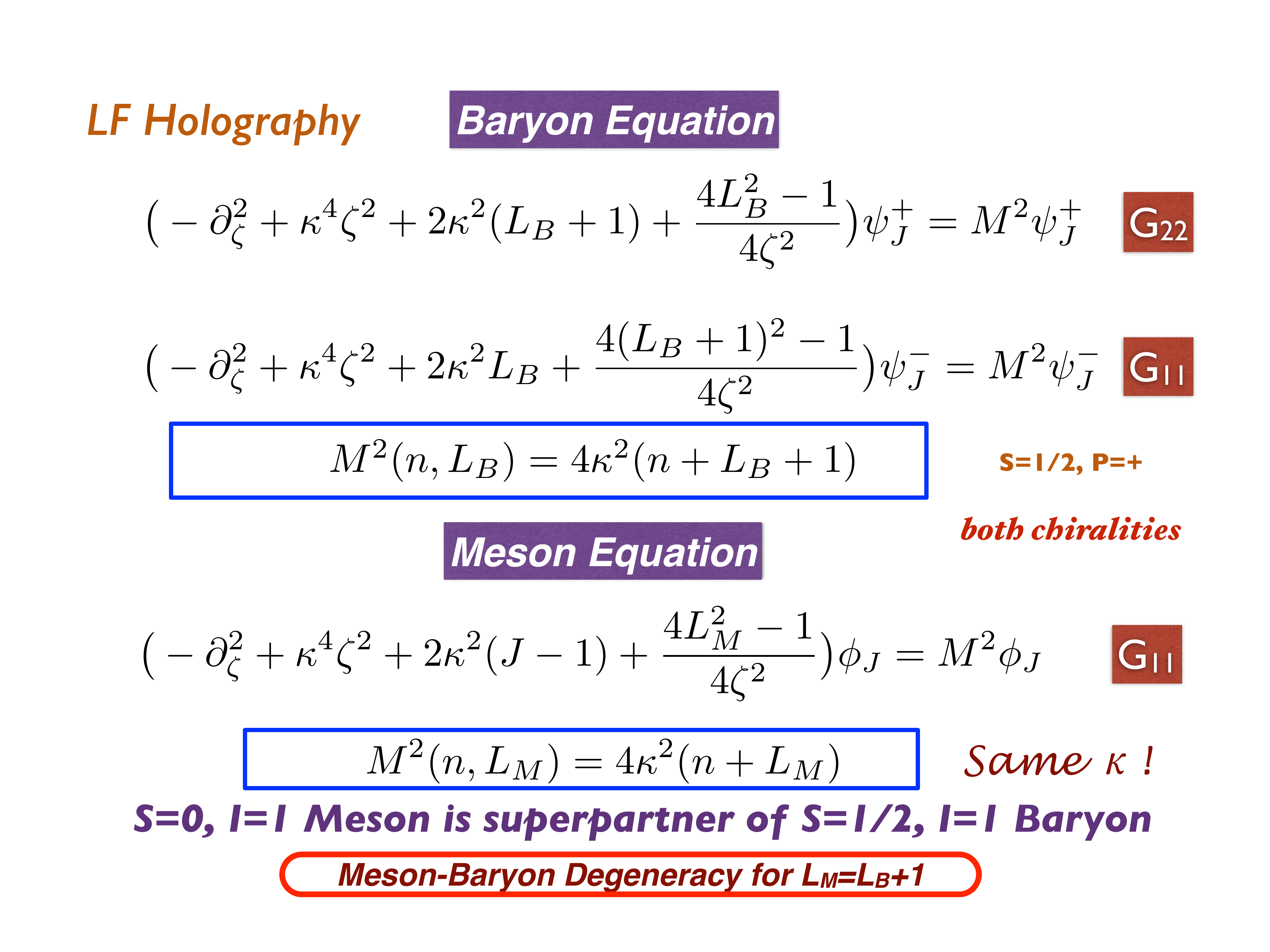}
\includegraphics[height=10cm,width=15cm]{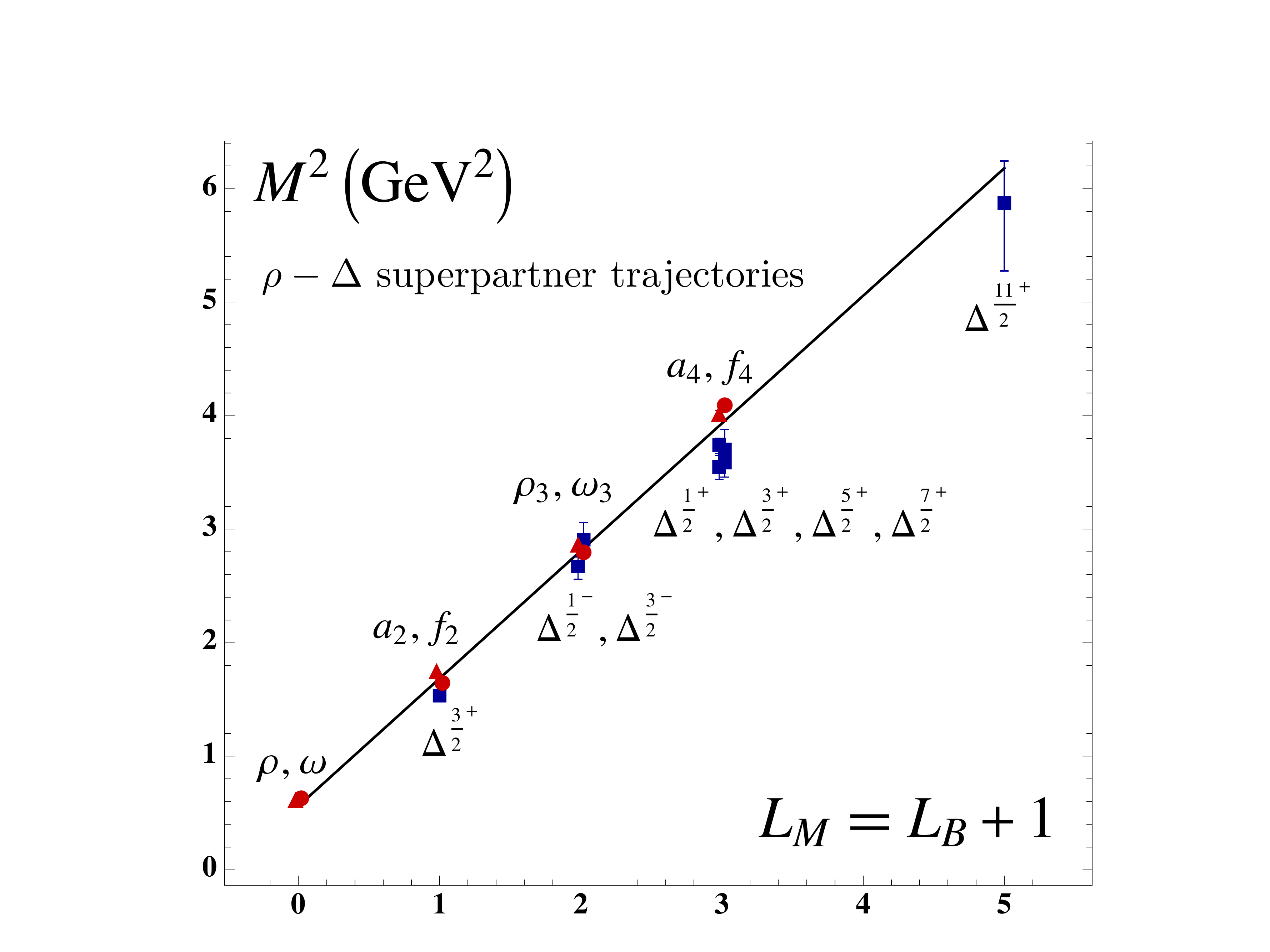}
\end{center}
\caption{(A). The LF Schr\"odinger equations for baryons and mesons for zero quark mass derived from the Pauli $2\times 2$ matrix representation of superconformal algebra.  
The $\psi^\pm$  are the baryon quark-diquark LFWFs where the quark spin $S^z_q=\pm 1/2$ is parallel or antiparallel to the baryon spin $J^z=\pm 1/2$.   The meson and baryon equations are identical if one identifies a meson with internal orbital angular momentum $L_M$ with its superpartner baryon with $L_B = L_M-1.$
See Refs.~\cite{deTeramond:2014asa,Dosch:2015nwa,Dosch:2015bca}.
(B). Comparison of the $\rho/\omega$ meson Regge trajectory with the $J=3/2$ $\Delta$  baryon trajectory.   Superconformal algebra  predicts the degeneracy of the  meson and baryon trajectories if one identifies a meson with internal orbital angular momentum $L_M$ with its superpartner baryon with $L_M = L_B+1.$
See Refs.~\cite{deTeramond:2014asa,Dosch:2015nwa}.}
\label{FigsJlabProcFig3.pdf}
\end{figure}

The supercharges connect the baryon and meson spectra  and their Regge trajectories to each other in a remarkable manner: each meson has internal  angular momentum one unit higher than its superpartner baryon  $L_M = L_B+1.$  See  Fig. \ref {FigsJlabProcFig3.pdf}(A).   Only one mass parameter $\kappa$ again appears; it sets the confinement and the hadron mass scale in the  chiral limit, as well as  the length scale which underlies hadron structure.  ``Light-Front Holography"  not only predicts meson and baryon  spectroscopy  successfully, but also hadron dynamics, including  vector meson electroproduction,  hadronic light-front wavefunctions, distribution amplitudes, form factors, and valence structure functions.  
The LF Schr\"odinger Equations for baryons and mesons derived from superconformal algebra  are shown  in Fig. \ref{FigsJlabProcFig3.pdf}.
In effect the baryons on the proton (Delta) trajectory are bound states of a quark with color $3_C$ and scalar (vector)  diquark with color $\bar 3_C$ 
The proton eigenstate labeled $\psi^+$ (parallel quark and baryon spins) and $\psi^-$ (anti parallel quark and baryon spins)  have equal Fock state probability -- a remarkable feature of chirality invariance.  The static properties of the nucleons is discussed in ref. ~\cite{Liu:2015jna}

The comparison between the meson and baryon masses of the $\rho/\omega$ Regge trajectory with the spin-$3/2$ $\Delta$ trajectory 
is shown in Fig. \ref{FigsJlabProcFig3.pdf}(B).
Superconformal algebra  predicts that the bosonic meson and fermionic baryon masses are equal if one identifies each meson with internal orbital angular momentum $L_M$ with its superpartner baryon with $L_B = L_M-1.$      Since $|L_B - L_M | =1,$    The meson and baryon  superpartners thus have have the same parity as well as the same twist.

Notice that the twist  $ 2+ L_M = 3 + L_B$ of the interpolating operators for the meson and baryon superpartners are the same.   Superconformal algebra also predicts that the LFWFs of the superpartners are identical, and thus they have identical LFWFs, and the corresponding elastic and transition form factors are equal.   The predicted identity of meson and baryon timelike form factors can be tested in $e^+ e^- \to H \bar H^\prime $ reactions. 

In the case of $e p \to e^\prime X$, one can consider the collisions of the confining  QCD flux tube appearing between the $q$ and $\bar q$  of the virtual photon with the flux tube between the quark and diquark of the proton.   Since the $q \bar q $ plane is aligned with the scattered electron's plane, the resulting ``ridge"  of hadronic multiplicity produced from the $\gamma^* p$ collision will also tend to be aligned with the scattering plane of the scattered electron.  The virtual photon's flux tube will also depend on the photon virtuality $Q^2$, as well as the flavor of the produced pair arising from $\gamma^* \to q \bar q$.  In the case of high energy $\gamma^* \gamma^*$ collisions, one can control the  produced  hadron multiplicity and ridge geometry using the scattered electrons' planes. The resulting dynamics~\cite{Brodsky:2014hia} is  a natural extension of the flux-tube collision description of the ridge produced in $p-p$ collisions~\cite{Bjorken:2013boa}.

\section {The QCD Coupling at all Scales} 

The QCD running coupling $\alpha_s(Q^2)$
sets the strength of  the interactions of quarks and gluons as a function of the momentum transfer $Q$.
The dependence of the coupling
$Q^2$ is needed to describe hadronic interactions at 
both long and short distances. 
The QCD running coupling can be defined~\cite{Grunberg:1980ja} at all momentum scales from a perturbatively calculable observable, such as the coupling $\alpha^s_{g_1}(Q^2)$, which is defined from measurements of the Bjorken sum rule.   At high momentum transfer, such ``effective charges"  satisfy asymptotic freedom, obey the usual pQCD renormalization group equations, and can be related to each other without scale ambiguity 
by commensurate scale relations~\cite{Brodsky:1994eh}.  

\begin{figure}
\begin{center}
\includegraphics[height=10cm,width=15cm]{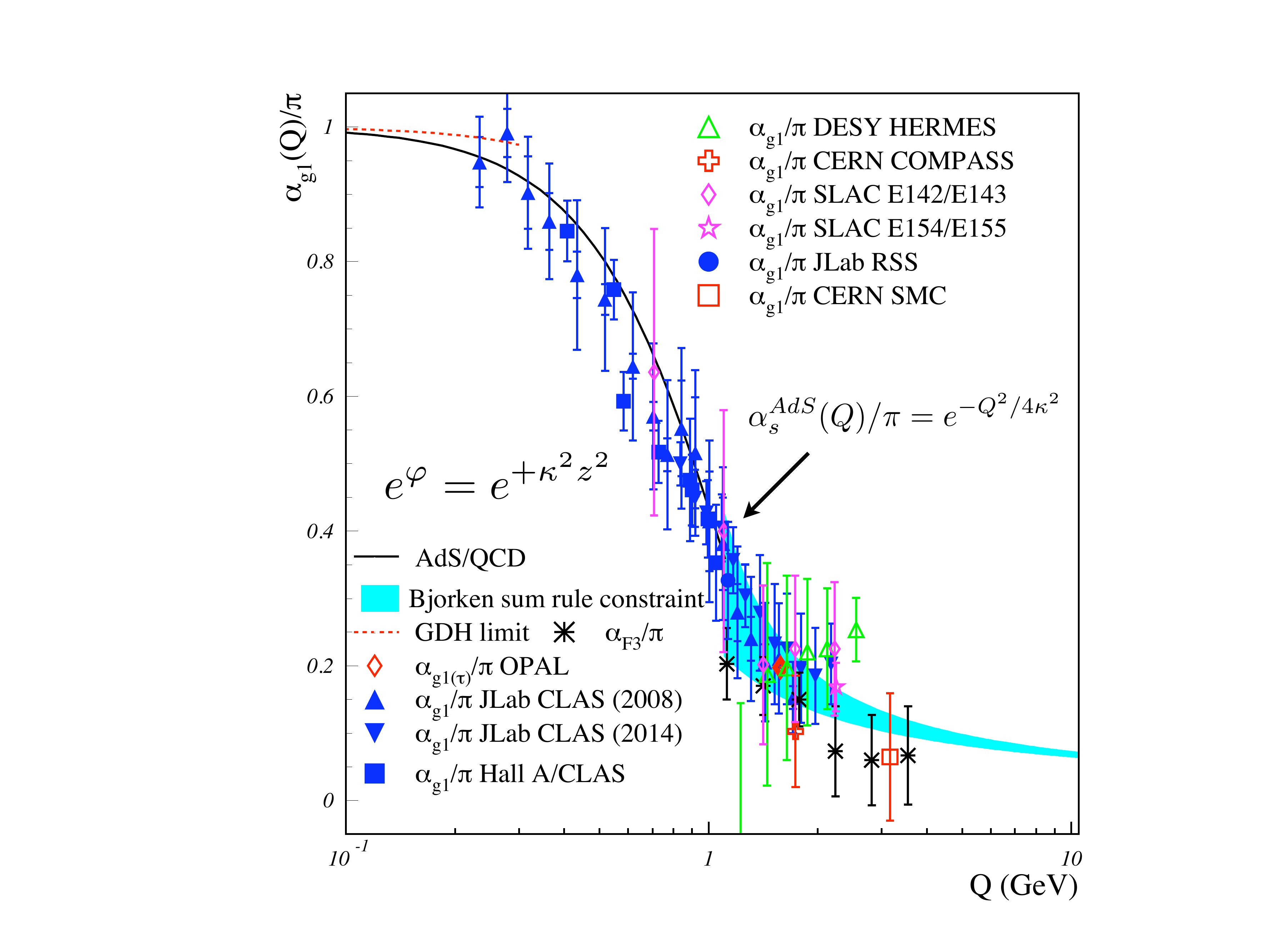}
\includegraphics[height=10cm,width=15cm]{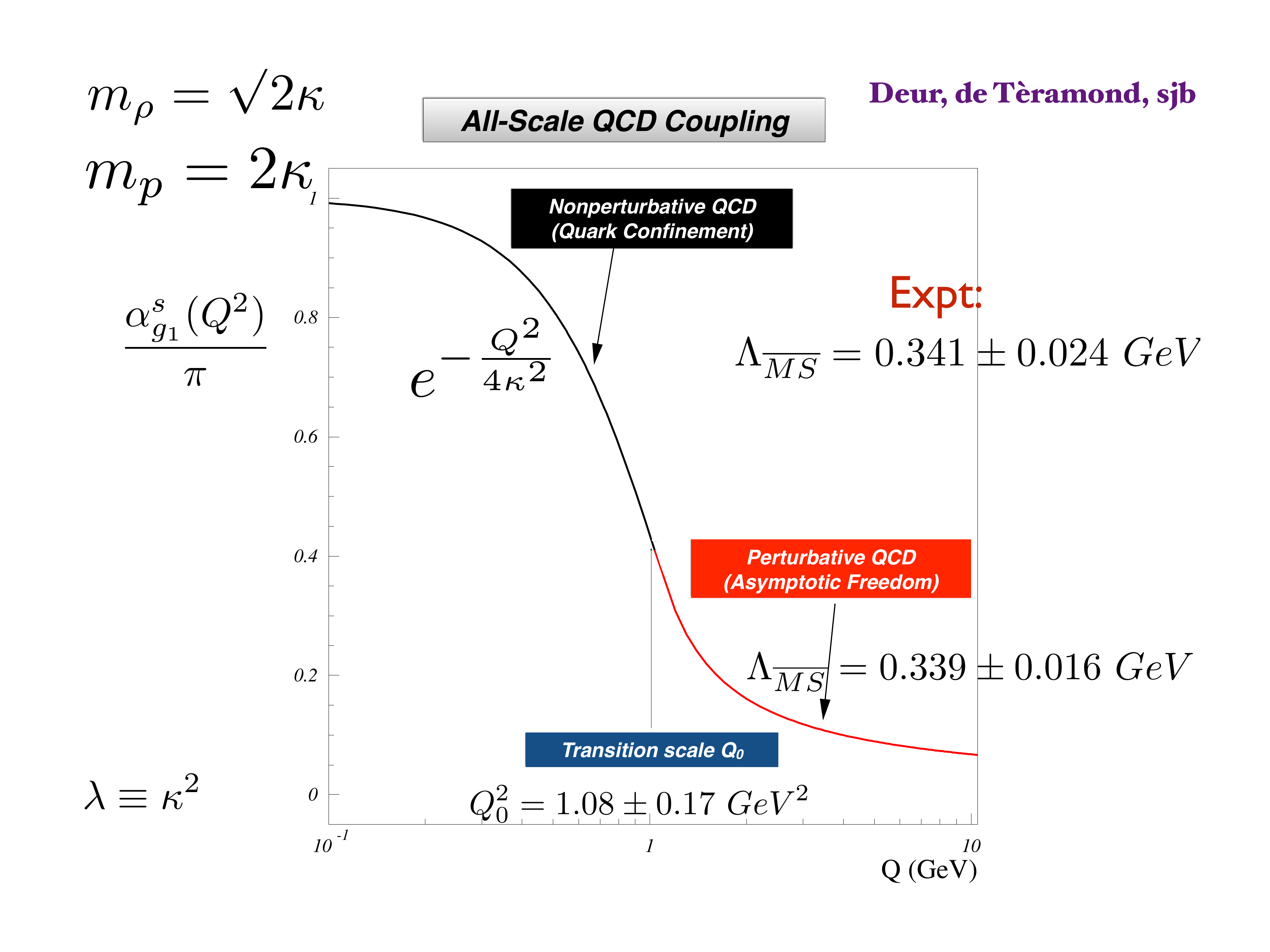}
\end{center}
\caption{
(A) Comparison of the predicted nonpertubative coupling, based on  the dilaton $\exp{(+\kappa^2 z^2)}$ modification of the AdS$_5$ metric, with measurements of the effective charge $\alpha^s_{g_1}(Q^2)$  
defined from the Bjorken sum rule.
(B)  Prediction from LF Holography and pQCD for the QCD running coupling $\alpha^s_{g_1}(Q^2)$ at all scales.   The magnitude and derivative of the perturbative and nonperturbative coupling are matched at the scale $Q_0$.  This matching connects the perturbative scale 
$\Lambda_{\overline{MS}}$ to the nonpertubative scale $\kappa$ which underlies the hadron mass scale. 
See Ref.~\cite{Brodsky:2014jia}. 
}
\label{FigsJlabProcFig5.pdf}
\end{figure}

The dilaton  $e^{+\kappa^2 z^2}$ soft-wall modification of the AdS$_5$ metric, together with LF holography, predicts the functional behavior of the running coupling
in the small $Q^2$ domain~\cite{Brodsky:2010ur}: 
${\alpha^s_{g_1}(Q^2) = 
\pi   e^{- Q^2 /4 \kappa^2 }}. $ 
Measurements of  $\alpha^s_{g_1}(Q^2)$ are remarkably consistent~\cite{Deur:2005cf}  with this predicted Gaussian form; the best fit gives $\kappa= 0.513 \pm 0.007~GeV$.   
See Fig.~\ref{FigsJlabProcFig5.pdf}(A)

Deur, de Teramond, and I~\cite{Brodsky:2010ur,Deur:2014qfa,Brodsky:2014jia} have shown how the parameter $\kappa$,  which   determines the mass scale of  hadrons in the zero quark mass limit, can be connected to the  mass scale $\Lambda_s$  controlling the evolution of the perturbative QCD coupling.  The high momentum transfer dependence  of the coupling $\alpha_{g1}(Q^2)$ is  specified by  pQCD and its renormalization group equation.  The 
matching of the high and low momentum transfer regimes  of $\alpha_{g1}(Q^2)$ -- both its value and its slope -- then determines the scale $Q_0$ setting the interface between perturbative and nonperturbative hadron dynamics.  This connection can be done for any choice of renormalization scheme, such as the $\overline{MS}$ scheme,
as seen in  Fig.~\ref{FigsJlabProcFig5.pdf} (B).   
The result of this perturbative/nonperturbative matching is an effective QCD coupling  defined at all momenta.   

The predicted value of $\Lambda_{\overline{MS}} = 0.341 \pm 0.024~GeV$ from this analysis agrees well the measured value~\cite{Agashe:2014kda}  
$\Lambda_{\overline{MS}} = 0.339 \pm 0.016~GeV.$
These results, combined with the AdS/QCD superconformal predictions for hadron spectroscopy, allow us to compute hadron masses in terms of $\Lambda_{\overline{MS}}$:
$m_p =  \sqrt 2 \kappa = 3.21~ \Lambda_{\overline{MS}},~ m_\rho = \kappa = 2.2 ~ \Lambda_{\overline{ MS} }, $ and $m_p = \sqrt 2 m_\rho, $ meeting a challenge proposed by Zee~\cite{Zee:2003mt}.
The pion is predicted to be massless for $m_q=0$  consistent with chiral theory.

The value of $Q_0$ can be used to set the factorization scale for DGLAP evolution of hadronic structure functions and the ERBL evolution of distribution amplitudes.
Deur, de T\'eramond, and I~\cite{Deur}, have also computed the dependence of $Q_0$ on the choice of the  effective charge used to define the running coupling and the renormalization scheme used to compute its behavior in the perturbative regime.

\section{Hadronization at the Amplitude Level and other New Directions}

\begin{itemize}

\item
The new insights into color confinement given by AdS/QCD suggest that one could compute hadronization at  amplitude level~\cite{Brodsky:2009dr} using LF time-ordered perturbation theory, but including the confinement interaction.  For example, if one computes $e^+ e^- \to q \bar q \to q \bar q g \cdots$, the quarks and gluons only appear in intermediate states, and only hadrons can be produced.  LF perturbation theory 
provides a remarkably efficient method for the calculation of multi-gluon amplitudes~\cite{Cruz-Santiago:2015nxa}. 

\item
The eigensolutions of the AdS/QCD LF Hamiltonian can used to form an ortho-normal basis for diagonalizing the complete QCD LF Hamiltonian.  This method, ``basis light-front quantization"~\cite{Vary:2009gt}  is expected to be more efficient than the DLCQ method~\cite{Pauli:1985pv} for obtaining QCD 3+1  solutions.

\item
All of the hadron physics predictions discussed in this report are independent of the value of $\kappa$; only dimensionless ratios are predicted, such as $m_p = \sqrt 2 m_\rho$ and the ratio $\Lambda_{\overline  MS}/m_\rho$.    The ratio can be obtained in any renormalization scheme. 
One thus retains dilatation invariance   $\kappa \to \gamma  \kappa$ of the prediction..

\item
The $\kappa^4 \zeta^2$ confinement interaction between a $q$ and $\bar q$ will induce a $\kappa^4/s^2$ correction to $R_{e^+ e^-}$, replacing the $1/ s^2$ signal usually attributed to a vacuum gluon condensate.  

\item
The kinematic condition that all $k^+ = k^0+ k^3$ are positive and conserved precludes QCD condensate contributions to the $P^+=0$ LF vacuum state, which by definition is the causal, frame-independent lowest invariant mass eigenstate of the LF Hamiltonian~\cite{Brodsky:2009zd,Brodsky:2012ku}. 

\item
It is interesting to note that the contribution of the {\it `H'} diagram to $Q \bar Q $ scattering is IR divergent as the transverse separation between the $Q$  
and the $\bar Q$ increases~\cite{Smirnov:2009fh}.  This is a signal that pQCD is inconsistent without color confinement.  The sum of such diagrams could sum to the confinement potential $\kappa^4 \zeta^2 $ dictated by the dAFF principle that the action remains conformally invariant despite the mass scale in the Hamiltonian.

\end{itemize}

\section{Elimination of  Renormalization and Factorization Scale Ambiguities}

The ``Principle of Maximum Conformality", (PMC)~\cite{Wu:2013ei} systematically eliminates the renormalization scale ambiguity in perturbative QCD calculations, order-by-order.    The PMC predictions are also insensitive to the choice of the initial renormalization scale $\mu_0.$
The PMC sums all of the non-conformal terms associated with the QCD $\beta$ function into the scales of the coupling at each order in pQCD, systematically extending the BLM procedure~\cite{Brodsky:1982gc} to all orders. 
The resulting  conformal series is free of renormalon resummation problems.  The number
of active flavors $n_f$ in the QCD $\beta$ function is also
correctly determined at each order. 
 
The $R_\delta$ scheme -- a generalization  of t'Hooft's  dimensional regularization systematically  identifies the nonconformal $\beta$ contributions to any perturbative QCD series, thus allowing the automatic implementation of the PMC procedure~\cite{Mojaza:2012mf}.     
 The resulting scale-fixed predictions for physical observables using
the PMC are {\it  independent of
the choice of renormalization scheme} --  a key requirement of 
renormalization group invariance.    
A related approach is given in Refs.~\cite{Kataev:2014zha,Kataev:2014jba,Kataev:2014zwa}.
 
The elimination of renormalization scale ambiguities greatly increases the precision, convergence, and reliability of pQCD predictions.  
For example, PMC scale-setting has been applied to the pQCD prediction for $t \bar t$ pair production at the LHC,  where subtle aspects of the renormalization scale of the three-gluon vertex and multi-gluon amplitudes, as well as  large radiative corrections to heavy quarks at threshold play a crucial role.  
The large discrepancy of pQCD predictions with  the $t \bar t$  forward-backward asymmetry measured at the Tevatron is significantly reduced from 
$3~\sigma$ to approximately $ 1~\sigma$~\cite{Brodsky:2012rj,Brodsky:2012sz,Wu:2015rga,Wang:2015lna}.  

The use of  the scale $Q_0$ discussed in the previous section to  resolve  the factorization scale uncertainty in structure functions and fragmentation functions,  in combination with the PMC for  setting the  renormalization scales,  can 
greatly improve the precision of pQCD predictions for collider phenomenology.

\section{Is the Momentum Sum Rule Valid for Nuclear Structure Functions? }

Sum rules for deep inelastic scattering are usually analyzed using the operator product expansion of the forward virtual Compton amplitude, assuming it depends in the limit $Q^2 \to \infty$ on matrix elements of local operators such as the energy-momentum tensor.  The moments of structure functions and other distributions can then be evaluated as overlaps of the target hadron's light-front wavefunction,  as in the Drell-Yan-West formulae for hadronic form factors~\cite{Brodsky:1980zm,Liuti:2013cna,Mondal:2015uha,Lorce:2011dv}.
The real phase of the resulting DIS amplitude and its OPE matrix elements reflects the real phase of the stable target hadron's wavefunction.

The ``handbag" approximation to deeply virtual Compton scattering also defines the ``static"  contribution~\cite{Brodsky:2008xe,Brodsky:2009dv} to the measured parton distribution functions (PDF), transverse momentum distributions, etc.  The resulting momentum, spin and other sum rules reflect the properties of the hadron's light-front wavefunction.
However, final-state interactions which occur {\it after}  the lepton scatters on the quark, can give non-trivial contributions to deep inelastic scattering processes at leading twist and thus survive at high $Q^2$ and high $W^2 = (q+p)^2.$  For example, the pseudo-$T$-odd Sivers effect~\cite{Brodsky:2002cx} is directly sensitive to the rescattering of the struck quark. 
Similarly, diffractive deep inelastic scattering (DDIS)  involves the exchange of a gluon after the quark has been struck by the lepton~\cite{Brodsky:2002ue}.  In each case the corresponding DVCS amplitude is not given by the handbag diagram since interactions between the two currents are essential.
These ``lensing" corrections survive when both $W^2$ and $Q^2$ are large since the vector gluon couplings grow with energy.  Part of the final state phase can be associated with a Wilson line as an augmented LFWF~\cite{Brodsky:2010vs} which does not affect the moments.  

The Glauber propagation  of the vector system $V$ produced by the  DDIS interaction on the nuclear front face and its subsequent  inelastic interaction with the nucleons in the nuclear interior $V + N_b \to X$ occurs after the lepton interacts with the struck quark.  
Because of the rescattering dynamics, the DDIS amplitude acquires a complex phase from Pomeron and Regge exchange;  thus final-state  rescattering corrections lead to  nontrivial ``dynamical" contributions to the measured PDFs; i.e., they involve physicL aspects of the scattering process itself~\cite{Brodsky:2013oya}.  The $ I = 1$ Reggeon contribution to DDIS on the front-face nucleon leads to flavor-dependent antishadowing~\cite{Brodsky:1989qz,Brodsky:2004qa}.  This could explain why the NuTeV charged current measurement $\mu A \to \nu X$ scattering does not appear to show antishadowing
 in contrast to deep inelastic electron nucleus scattering as discussed in ref. ~\cite{Schienbein:2007fs}.
Again the corresponding DVCS amplitude is not given by the handbag diagram since interactions between the two currents are essential.

Diffractive deep inelastic scattering is leading-twist and is the essential component of the two-step amplitude which causes shadowing and antishadowing of the nuclear PDF.  It is important to analyze whether the momentum and other sum rules derived from the OPE expansion in terms of local operators remain valid when these dynamical rescattering corrections to the nuclear PDF are included.   The OPE is derived assuming that the LF time separation between the virtual photons in the forward virtual Compton amplitude 
$\gamma^* A \to \gamma^* A$  scales as $1/Q^2$.
However, the propagation  of the vector system $V$ produced by the DDIS interaction on the front face and its inelastic interaction with the nucleons in the nuclear interior $V + N_b \to X$ are characterized by a longer LF time  which scales as $ {1/W^2}$.  Thus the leading-twist multi-nucleon processes that produce shadowing and antishadowing in a nucleus are evidently not present in the $Q^2 \to \infty$ OPE analysis.

It should be emphasized  that shadowing in deep inelastic lepton scattering on a nucleus  involves  nucleons at or near the front surface; i.e, the nucleons facing the incoming lepton beam. This  geometrical orientation is not built into the frame-independent nuclear LFWFs used to evaluate the matrix elements of local currents.  Thus the dynamical phenomena of leading-twist shadowing and antishadowing appear to invalidate the sum rules for nuclear PDFs.  The same complications occur in the leading-twist analysis of deeply virtual Compton scattering $\gamma^* A \to \gamma^* A$ on a nuclear target.

\section*{Acknowledgments}

Presented at Light-Cone 2015, {\it Theory and Experiment for Hadrons on the Light-Front}, September 21-25, 2015,  INFN, Frascati, Italy. 
The results presented here are based on collaborations  and discussions  with  
James Bjorken, Kelly Chiu, Alexandre Deur, Guy de T\'eramond, Guenter Dosch, Susan Gardner, Fred Goldhaber,  Paul Hoyer, Dae Sung Hwang,  Rich Lebed, 
Simonetta Liuti, Cedric Lorce, Matin Mojaza,  Michael Peskin, Craig Roberts, Robert Shrock, Ivan Schmidt, Peter Tandy, and Xing-Gang Wu.
This research was supported by the Department of Energy,  contract DE--AC02--76SF00515.  
SLAC-PUB-16453.

\end{document}